\newcommand{\un}{~\mathrm}

\documentclass{elsart}

\usepackage[T1]{fontenc}
\usepackage{graphicx}
\usepackage{dcolumn}
\usepackage{bm}
\usepackage{amssymb}

\begin{document}

\begin{frontmatter}



\title{Crack path instabilities in DCDC experiments in the low speed regime.}

\author {C. Marlière\corauthref{cor1}},
\author {F. Despetis} and
\author {J. Phalippou}
\author {}
\address {Laboratoire des Verres - UMR CNRS-UM2 5587, Université Montpellier 2, C.C. 69 - Place Bataillon, F-34095 Montpellier Cedex 5 - France}

\corauth[cor1]{Phone: +33 4 67 14 37 06, Fax: +33 4 67 14 34 98}

\ead{marliere@univ-montp2.fr (C. Marlière)} 

\begin{abstract}

We studied the low speed fracture regime ($10^{-4}\un{m.s}^{-1}$ - $10^{-9}\un{m.s}^{-1}$) in different glassy materials (soda-lime glass, glass-ceramics) with variable but controlled length scale of heterogeneity. The chosen mechanical system enabled us to work in pure mode I (tensile) and at a fixed {\em load} on DCDC (double cleavage drilled compression) specimen. The internal residual stresses of studied samples were carefully relaxed by appropriate thermal treatment. By means of optical and atomic force (AFM) microscopy techniques fracture surfaces have been examined. We have shown for the first time that the crack front line underwent an out-of-plane oscillating behavior as a result of a reproducible sequence of instabilities. The wavelength of such a phenomenon is in the micrometer range and its amplitude in the nanometer range. These features were observed for different glassy materials providing that a typical length scale characterizing internal heterogeneities was lower than a threshold limit estimated to few nanometers. This effect is the first clear experimental evidence of crack path instabilities in the low speed regime in a uniaxial loading experiment. This phenomenon has been interpreted by referring to the stability criterion for a straight crack propagation as presented by Adda-Bedia {\em et al}. \cite{ref1}.

\end{abstract}

\end{frontmatter}


\section{Introduction}

Quasistatic brittle fracture is a subject of importance for both practical and fundamental reasons. Two classes of experimental set-up have been primarily developed in order to propagate cracks in a stable and controllable manner. The first one is based on a thermally induced stress field \cite{ref2,ref3,ref4}. An other classical way to propagate slow cracks is to use double cleavage drilled compression (DCDC) specimen. This method initially developed by Janssen \cite{ref5} has numerous advantages: compression loading, mid plane crack stability, auto precracking and pure mode I (tensile) crack tip opening. 
However numerous examples of unstable straight crack propagation were evidenced in thermal experiments even in the low speed regime. According to values of thermal gradient and width of the glass plate, three types of crack propagation regime can develop from the small initial notch made to ensure the nucleation of a single propagating crack : no propagation, straight propagation and wavy propagation. Such propagating cracks were observed for velocities between $10^{-2}\un{m.s}^{-1}$ and $10^{-5}\un{m.s}^{-1}$. An attempt to explain this behavior was put forward \cite{ref2} using the Cotterell and Rice criterion \cite{ref6} but partially failed when the real experimental temperature profile was measured \cite{ref3}. More recently the concept of maximum tangential stress criterion, first introduced by Erdogan {\em et al.} \cite{ref7}, was used to explain a complete set of thermal experiments \cite{ref4}.
Such unstable behavior in quasi-static crack propagation in uniaxial loading conditions was predicted by Adda-Bedia {\em et al. } \cite{ref1} but was surprisingly never experimentally observed, probably because of the extreme sensitivity of this phenomenon to experimental conditions \cite{ref1}.
By working at a fixed {\em load} on DCDC specimen, the internal residual stresses of which were carefully relaxed by appropriate thermal treatment, we have shown - for the first time to our knowledge -  that the crack front line undergoes an oscillating behavior as a result of a reproducible sequence of instabilities. The wavelength of this wavy behavior is in the micrometer range and its amplitude in the nanometer one. 

\section{Experimental }

DCDC samples were parallelepipedic ($4\times4\times40\un{mm}^3$) with a central hole of 1mm in diameter (nominal) drilled perpendicularly to two of the large parallel faces. Details of experiments have been reported in a previous paper \cite{ref8}. The sample was then placed, vertically, on its smallest face, in equilibrium on the horizontal basis of our experimental set-up. By means of a freely moving piston, the upper face of the specimen was then loaded by an increasing weight at a rate of about 25N.$\un{s}^{-1}$  till the crack was initiated. Then the weight on the sample was maintained constant at a value of typically 1.5kN. These experiments were done at a constant temperature of $22.0\pm 1^\circ\mathrm{C}$ and at a relative humidity of $50\pm 5\%$. 

The crack front position was measured in the middle of the specimen (at equal distances of the lateral surfaces) by i) a standard video-recorder system for crack speed between $10^{-4}\un{m.s}^{-1}$ and $10^{-6}\un{m.s}^{-1}$ and ii) an optical system with a magnification of $\times $40 for lower crack speeds ($10^{-6}\un{m.s}^{-1}$ - $10^{-9}\un{m.s}^{-1}$). In the latter case, the position of crack front relatively to the center of the specimen was measured with an accuracy of 10$\un{\mu}$m. The stress intensity factor, $K_I$, as well as the mean crack speed, $v$, was deduced from the measured crack length, using reference \cite{ref9}. The $v(K_I)$ curve were then plotted. When the crack speed reached a value lower than $10^{-9}\un{m.s}^{-1}$ the samples were reloaded in order to gently fracture the sample all over its length. Special care was taken in order to preserve the two halves of the specimen from subsequent damage. 

The fracture surfaces were analyzed by an experimental system (Veeco, D3100) combining optical microscopy and atomic force microscopy (AFM). Both methods could be simultaneously performed. The samples were glued on the sample holder. This holder was moveable under the fixed AFM-optical system using a stepper motor. The position of the center of the studied portion of surface relatively to the center of the specimen was measured with an accuracy of less than $10\un{\mu m}$. In this manner, it was possible to identify for every point of the AFM or optical images the value of $K_I$ and the speed the crack had when it passed through this point. 

AFM experiments were done in a high amplitude resonant mode ("tapping" mode). Experiments were performed in ambient conditions at a relative humidity of $35\pm 5\%$ and a temperature of $22.0\pm 0.5~^\circ\mathrm{C}$. More details are given in reference \cite{ref10}.

Experiments were performed on two kinds of materials. The first one was soda-lime silicate glass. A thermal treatment ($530^\circ\mathrm{C}$) was done before fracture experiment in order to remove residual stresses. The second set of samples was made from a lithium alumino-silicate glass-ceramics. By modifying thermal treatments three types of structure were observed : the first one (sample A) corresponds to a pure glassy state ($660^\circ\mathrm{C}$) as identified by X-ray diffraction (Fig.~\ref{fig1}a). The second one (sample B) is classified as a slightly unglassy state in which small crystals of $\beta $-quartz phase have been nucleated (Fig.~\ref{fig1}b). This structure was obtained after a two-step thermal treatment at plateau temperatures of $T_{1}$=$750~^\circ\mathrm{C}$  and $T_{2}$=$900~^\circ\mathrm{C}$ . Samples C exhibits further crystallization (increased number of crystallites of larger size) due to a higher temperature ($950~^\circ\mathrm{C}$) for the second step of thermal treatment (Fig.~\ref{fig1}c).

\section{Results}

In the Figure~\ref{fig2} are plotted typical $v(K_I)$ curves as obtained with the different types of samples we studied. A noticeable feature is that for the glass ceramics (open symbols) the curves are shifted towards higher $K_I$  values when the thermal treatment is performed. The reproducibility of these experiments is revealed  by the very good superimposition of the two $v(K_I)$ curves performed on two different samples (B1 and B2: up and down triangles) which underwent the same thermal treatment. 
The fracture surfaces were then systematically characterized by our optical-AFM system. We put a special attention to the part of the surface fracture  corresponding to the low speed regime ($10^{-4}\un{m.s}^{-1}$ - $10^{-9}\un{m.s}^{-1}$). 
We firstly put the stress on the lowest speed regime ($v$  < $10^{-8}\un{m.s}^{-1}$). Scanning Electron Microsocpy (SEM) experiments revealed that corresponding fracture surfaces became increasingly rougher with harsher thermal treatments -from sample A (Fig.~\ref{fig3}a) to sample C (Fig.~\ref{fig3}c). These variations of  RMS roughness were quantified by AFM measurements on 10$\un{\mu}$m$\times $10$\un{\mu}$m surfaces where crack speed was in the range of few $10^{-9}\un{m.s}^{-1}$. The results are reported on  Table \ref{tableau1}.
It must be emphasized that RMS roughness values for pure glassy materials are very similar whatever the chemical composition of glasses. For the glass ceramics with an increasing number of larger $\beta $-quartz nanocrystals, however, an important enhancement of the RMS roughness is observed. The values of RMS roughness are thus directly related to the length scale of structural heterogeneities. 

For higher crack speeds (except for glass-ceramics sample C; see below) an original phenomenon was observed (Fig.~\ref{fig4}). Fracture surface observation revealed the presence of periodic variations in the crack path (in the direction normal to the mean crack plane) along the crack propagation direction. These corrugations in the crack path were observed : i) all along the virtual lines followed by both crack fronts propagating inside the specimen and symmetrically about the central hole, ii) on both fracture surfaces. An example (soda-lime glass, average crack speed $v\cong$  2.$10^{-5}\un{m.s}^{-1}$  and $K_I \cong\un{0.6MPa.m}^{1/2}$ ) is shown on Figure~\ref{fig4}. The 70$\un{\mu}$m$\un{\times}$70$\un{\mu}$m AFM scan of Figure~\ref{fig4}a shows alternate black (deep) and white (high) bands parallel to the crack front. The sine-like shape of these roughness oscillations is better evidenced in Figure~\ref{fig4}c where all the lines parallel to the crack propagation direction (i.e. perpendicular to the black and white bands) of an independent 10$\un{\mu}$m$\un{\times}$10$\un{\mu}$m  AFM scan (Figure ~\ref{fig4}b) are averaged. 
The occurrence (or not) of such a phenomenon was monitored for all the previously described samples. The following common features were found: 
\begin {itemize} 
\item[$\bullet$ ] Samples of materials in pure glassy state (soda-lime and glass ceramics) revealed this wavy behavior for a large range of $K_I$ values. For instance, in the case of soda-lime glass, the $K_I$ domain extended between the following values: $K_I\cong\un{0.5 MPa.m}^{1/2}$ and $K_I\cong\un{0.8 MPa.m}^{1/2}$   respectively corresponding to crack front speeds of 1.5 $10^{-6}\un{m.s}^{-1}$ and $10^{-4}\un{m.s}^{-1}$. Between these two limiting values the crack ran over a distance of 2 millimeters in approximately 5 minutes. However for samples of glass ceramics -in the pure glassy state- the oscillating domain is slightly narrower: $\Delta K_I\cong\un{0.17 MPa.m}^{1/2}$.
\item[$\bullet $ ] One striking feature is that the $K_I$-width of occurrence of the roughness waves is strongly reduced for samples in intermediate unglassy state (sample~B):  $\Delta K_I\cong\un{~0.08 MPa.m}^{1/2}$. Furthermore no such oscillations were observed on highly unglassy samples (sample C);
\item[$\bullet$ ]	The wavelength of these ripples, $\lambda $, is in the micrometer range and its scaling with $K_I$ values was systematically studied. Figure~\ref{fig5}  shows the typical variation of $\lambda $ as observed in the case of soda-lime samples. It is shown that a large decrease of the wavelength occurs with decreasing values of $K_I$. Similar variations -in the same range of wavelength- were observed for glassy or slightly unglassy samples.
\item[$\bullet$ ]	The amplitude of such corrugated oscillations is in the nanometer range.
\end  {itemize}


\section{Discussion}

The experimental study presented in this paper revealed that fracture surfaces of various glassy materials are modulated by roughness waves the wave vector of which is parallel to the crack propagation direction. These roughness waves were observed for the various materials studied for this paper providing that the length scale of their heterogeneities is lower than a threshold around 10~nm. This length scale of heterogeneities can be very roughly evaluated by the RMS roughness of surface created by crack propagating at low speed (lower than few $10^{-9}\un{m.s}^{-1}$), see Table \ref{tableau1}. It means that this phenomenon is directly correlated to the length scale of microstructural heterogeneities of materials used for these experiments.

It must be noted that a trivial AFM spurious effect can be ruled out because these ripples were observed too by optical microscopy for the largest wavelengths (largest $K_I$ values). It must be emphasized that, even if the aspect of these roughness ripples presents similarities with patterns observed in stress wave fractography, such an explanation is not relevant for the observations reported in this paper. Indeed no external transverse stress wave generator \cite{ref11,ref12}  was used. Furthermore the experimental set-up was working in a very quiet acoustical environment. Moreover the role of hypothetical residual external mechanical vibrations of the DCDC apparatus may be excluded. The resonance frequency and the quality factor of the equivalent mechanical system was indeed estimated to be higher than 70Hz and 6.$10^{4}$    respectively. Consequently the hypothetical residual vibrations would have a negligible amplitude at the frequency range (between 0.1Hz to 1Hz) compared to the observed roughness waves and calculated by the ratio between the crack front speed and the related wavelength. 

Similar observations of wavy behavior during crack tip propagation were previously reported \cite{ref2,ref4}. The steady propagation of a crack localized within a thermal gradient was studied and an oscillatory behavior for given values of crack speed and glass plate width, $W$, was observed. The transition between the regimes corresponding to linear crack propagation and wavy one was studied as a function of specimen width for fixed temperature jump. The experiments by Yuse {\em et al}. \cite{ref2}   revealed that both the wavelength and amplitude of oscillatory behavior induced by thermal stress were in the millimeter range. As these length scales were of the same order of magnitude as the width of the region of rapid temperature change, it was assumed \cite{ref2,ref4}  that the related instability mechanisms were only due to "external" parameters such as stresses induced by the thermal field and were independent of the microstructure of broken materials. Furthermore, Yuse's experiments \cite{ref2}  showed that the related wavelength scaled linearly with $W$. Referring to the study Marder \cite{ref13}, it is possible to plot the variation of wavelength, $\lambda$, versus $K_I$ for the thermal experiments (Fig.~\ref{fig6}). This plot reveals that $\un{log}(\lambda)$ slightly decreases with decreasing values of $K_I$. The data obtained in the present mechanical experiments can also be drawn using the same set of coordinates (see Fig.~\ref{fig6}). This plot shows that a similar behavior to that revealed by thermal experiments was observed for mechanical experiments but for slightly lower values of $K_I$. This suggests that this wavy behavior could originate from the same phenomenon. The main difference is that, in our case, small variations of $K_I$ cause larger changes in related roughness wavelength. 

A likely explanation for the occurrence of such oscillations in our DCDC experiments is that the crack front similarly undergoes a reproducible sequence of instabilities. Indeed such destabilizing effects in the case of uniaxial loading have already be predicted by Adda-Bedia {\em et al.} \cite{ref1}. These authors argued that, even in quasi-static conditions, crack propagation may become unstable: it may deviate from straight line in a direction perpendicular to the applied external uniaxial loading after which a wavy crack path may appear. As already mentioned in \cite{ref1}, stability thresholds drastically depend on both geometric and loading conditions. We think that our experimental set-up (DCDC experiments with a pure dead-weighted loading, carefully prepared samples) has been crucial to reveal such a phenomenon. 
As both wavelength and amplitude of the observed ripples are in the sub-micrometer range, the destabilizing factor -in the sense of Cotterell and Rice criterion \cite{ref6}- from a straight propagation of the crack likely has an "internal" origin as it is connected to the sub-micrometric structure of materials used for these experiments. Indeed, when the length scale of heterogeneity is low enough such wavy propagating behavior is observed. On the opposite, for materials with larger heterogeneities the fracture surface only presents an incoherent structure (with a much higher roughness). The length cutoff is then in the range of 10nm. 
The origin of this instability has likely to be found in the recently predicted corrugation waves \cite{ref14}  which can be generated when fracture occurs in mode I. These authors indeed explain that local distortion of the crack front, due for instance to the presence of structural heterogeneities, may give rise to non-planar deformations of the crack front, they called "corrugation waves". As this effect is assumed to be due to long-ranged elastic effects causing dynamic stress transfer along the crack front it likely explains why such phenomenon can be observed on long distances. It must be emphasized that these authors \cite{ref14} explain that the corrugation waves may occur even in the case of a very low mean crack speed. Indeed, as experimentally observed \cite{ref15,ref16}  the {\em local} crack speed may exceed the mean crack speed by several order of magnitudes and lead to avalanche-like processes. Furthermore the roughness wave amplitude as predicted in Ref. \cite{ref14}  for glassy materials ($\delta h \simeq \un{2nm}$) is in the same order of magnitude than that observed in the current study.

\section{Conclusion}

Experiments were done on a DCDC mechanical set-up working at a fixed {\em load}. The internal residual stresses of specimen in different glassy materials were carefully relaxed by appropriate thermal treatment. In the low speed regime ($10^{-4} - 10^{-9}\un{m.s}^{-1}$) we have demonstrated for the first time that the crack front line underwent an out-of-plane oscillating behavior along the direction of crack propagation. The wavelength was in the micrometer range and the peak-to-peak amplitude in the nanometer. This phenomenon was observed for different glassy materials providing that their typical length scale was lower than a threshold limit estimated to few nanometers. This effect was interpreted by the evidence of instabilities and long-range elastic effects occurring during the 'straight' propagation of the crack front. We believed that these instabilities are generated by structural heterogeneities on which crack front is pinned and depinned. Studies are now performed at sub-micrometer scale in order to better understand how this reproducible sequence of instabilities is developing.

\newpage

\begin{table}[!h]
\centering
\begin{tabular}{|l|r|}
\hline 
Soda-lime glass & 1~nm \\ 
\hline 
Glass ceramics : sample A & 1~nm \\
\hline 
Glass ceramics : sample B & 5~nm \\
\hline 
Glass ceramics : sample C & 10~nm \\
\hline
\end{tabular}
\caption{ RMS roughness on $\un{10 \mu m \times 10 \mu m}$ fracture surfaces (mean crack speed around few $10^{-9}\un{m.s}^{-1}$) for different substrates.} 
\label{tableau1}
\end{table}

\newpage

\begin{figure}
\centering
\caption{ X-ray diffraction patterns on glass-ceramics in a) pure glassy state, b) slightly unglassy state and c) more unglassy state.} 
\label{fig1}
\end{figure}

\newpage

\begin{figure}
\centering
\caption{ Crack speed versus stress intensity factor, $K_I$,  plots for different materials :
Full symbols : soda-lime glass;
Open symbols : Glass-ceramics (G.C.) 
$T_{1}$ and $T_{2}$ are the temperatures of the two-steps thermal treatment.
}
\label{fig2}
\end{figure}

\newpage

\begin{figure}
\centering
\caption{SEM pictures of fracture surfaces as generated by crack line propagation at an average mean speed in the $10^{-8}\un{m.s}^{-1}$ regime for glass-ceramics in a) pure glassy state, b) slightly unglassy state and c) more unglassy state. The crystalline phase was revealed by a mild HF etching (1\%). The white marker corresponds to 100nm.
 } 
\label{fig3}
\end{figure}

\newpage

\begin{figure}
\centering
\caption{AFM data of fracture surface of soda-lime glass; the vertical scale corresponds to 15 nm.
a) $\un{70 \mu m \times 70 \mu m}$  scan; b) $\un{10 \mu m \times 10 \mu m}$ scan; c) averaged profile of roughness oscillations as obtained on the $\un{10 \mu m \times 10 \mu m}$ scan.
}
\label{fig4}
\end{figure}

\newpage

\begin{figure}
\centering
\caption{Crack speed (squares) and roughness oscillations wavelength (circles) versus $K_I$ for soda-lime glass. The arrow indicates the case corresponding of AFM data of Figure 4.}
\label{fig5}
\end{figure}

\newpage

\begin{figure}
\centering
\caption{Wavelength of corrugation oscillations versus the stress intensity factor, $K_I$, for both thermal \cite{ref2} and mechanical (this paper) experiments. The dashed line is a guide for the eyes.
}
\label{fig6}
\end{figure}

\end{document}